# Thermal conductivity of free-standing silicon nanowire using Raman spectroscopy

Sandhyarani Sahoo,[1,2] Sameer Kumar Mallik,[1,2] Mousam Charan Sahu,[1,2] Anjana Joseph,[3] Bibhudutta Rout,[4] Gopal K. Pradhan,[5]* Satyaprakash Sahoo[1,2]*

[1]Laboratory for Low Dimensional Materials, Institute of Physics, Bhubaneswar, 751005, India

[2]Homi Bhabha National Institute, Training School Complex, Anushakti Nagar, Mumbai 400085, India

[3]Jawaharlal Nehru Centre for Advanced Scientific Research, Jakkur, Bangalore, 560064, India

[4]Department of Physics, University of North Texas, Denton, TX 76203, USA

[5]Department of Physics, School of Applied Sciences, KIIT Deemed to be University, Bhubaneswar, Odisha, 751024, India

**ABSTRACT:**

Low dimensional systems, nanowires, in particular, have exhibited excellent optical and electronic properties. Understanding the thermal properties in semiconductor nanowires is very important for their applications in their electronic devices. In the present study, the thermal conductivity of a freestanding silicon nanowire (NW) is estimated employing the Raman spectroscopy. The advantage of this technique is that the light source (laser) can be used both as heating and excitation source. The variations of the first-order Raman peak position of the freestanding silicon NW with respect to temperature and laser power are carried out. A critical analysis of effective laser power absorbed by exposed silicon NW, the detailed Raman study along with the concept of longitudinal heat distribution in silicon NW, the thermal conductivity of the freestanding silicon NW of 112 nm diameter is estimated to be ~53 W/m.K.

* Corresponding author: sahoo@iopb.res.in, gopal.pradhanfpy@kiit.ac.in

**INTRODUCTION:**

Silicon-based semiconductor technology has brought a technological revolution towards the progress of our society, which has benefited tremendously from its versatile use and remarkable capabilities. The development of electronics based on silicon and its supporting technology has brought down the physical size of devices and interconnects to the nanoscale regime. In this connection, NWs in general and silicon NWs, in particular, are promising materials for the miniaturization of devices towards the nanoscale regime.[1,2] Nowadays, the growth of NWs by different methods has been achieved with great control over their crystal quality, aspect ratio and heterostructures that has opened a new window to engineer various functional properties.[3–7] In a world of ever-shrinking circuits size due to the integration of various miniaturized components such as NWs, the management, and transport of heat is paramount. It is quite important for better understanding and designing heat transporting devices in a host of applications, from heat management in high-power electronics, batteries, and photovoltaics to thermoelectric energy harvesting and solid-state cooling. For example, applications such as thermoelectric energy harvesting, solid-state cooling, and thermal barrier coatings would require extremely low thermal conductivity. On the contrary, heat management in high-power electronics, batteries, and photovoltaics would require high thermal conductivity. Understanding and predicting the thermal conductivity of nanosystems will allow designing heat transport for a wide range of technological applications. Determination of the thermal conductivity of silicon NWs plays a pivotal role in two of their most fundamental potential applications: (i) heat dissipation for potential heat drain application, and (ii) thermoelectric materials. In these applications a combination of lower thermal conductivity and high electrical conductivity can result in NWs having superior performance than existing materials presently used for thermoelectric refrigeration.[8,9,10] A better understanding of thermal transport in nanostructures also has direct consequences in the design and performance of modern nanoelectronics for NW-based thermoelectric devices. Moreover, in the semiconducting NW system, the phonon contribution to the thermal conduction can be minimized by limiting the NW length scales smaller than phonon mean free path while keeping the contribution to the electrical conductivity unaltered.[11] Thus, it is quite clear that prior determination of the thermal conductivity of the NWs is of extreme importance.

Thermal conductivity in solids is due to the heat transfer by the electrons and phonons.[12] The predominant carriers of heat in the case of semiconductors and insulators are the phonons. Reduction of the dimension of the bulk material into nanostructure results in the spatial confinement of the thermal carriers. Movement of these carriers in a confined geometry will result in increased phonon boundary scattering and phonon–phonon scattering rate. Considering these facts, thermal conductivity as a function of silicon NW diameter has been investigated using molecular dynamics simulations by Ponomareva *et al.*[13] Their finding shows an interesting trend in thermal conductivity; predominant boundary scattering leads to suppressed thermal conductivity for NWs having diameter > 1.5 nm and phonon confinement effect results in enhanced thermal conductivity in silicon NWs having a diameter < 1.5nm.[13]

Measurement of the thermal conductivity of nanostructures such as graphene, carbon nanotubes, NWs, TMDCs is very challenging. In literature thermal conductivity of nanostructures is reported using micro heaters,[14] 3ω method,[15–17] Raman spectroscopy,[18–23] time-domain thermoreflectance (TDTR).[24] Among these techniques, Raman spectroscopy is found to be a simple and convenient technique to measure thermal conductivity has recently been applied to 2-D [18–21] as well as 1-D nanostructures.[22,23,25] The growing use of Raman thermometry arises from advantages lacking in other established techniques such as 3ω,[15,16] micro heating[14] that are relatively complex, require direct contact with sample with the necessity of microfabrication. Unlike the other standard optical techniques that measures the thermal conductivity only in the cross-plane direction, Raman technique measures the in-plane thermal conductivity. As it has been found in literature that for nanostructures having anisotropy in particular directions, the measurement of in-plane thermal conductivity is more relevant.[26] In Raman scattering method, the laser can be used both as heating and excitation source.[27–29] Analyzing the shift of the Raman peak position caused by uniform rise in temperature and laser power induced local heating, the thermal conductivity can be estimated.

Recently, measurements of the thermal conductivity of individual silicon NW is reported by Joule heating [14] or Raman thermometry method [23]. The Joule heating method is plagued by interfacial contact resistance as a potentially significant source of error, particularly with larger diameter NWs. In the second case, the NW was cantilevered.[23] In this work, we present a relatively simple method to measure the thermal conductivity of freestanding single silicon NW by using Raman spectroscopy. The first-order temperature coefficient ($\delta\omega/\delta T$) is obtained from the temperature-dependent Raman study and variation of Raman peak with laser power ($\delta\omega/\delta P$) is determined by power-dependent Raman study and further analysis is carried out to calculate thermal conductivity.

**EXPERIMENTAL:**

Silicon wafer containing silicon NWs is commercially procured from CVD Materials Corporation, USA has grown by VLS CVD using First NanoTM Easytube® system. Microstructure characterization of silicon NWs is performed using Field Emission Scanning Electron Microscope (Zeiss make). Raman spectra silicon NWs are recorded using a confocal micro-Raman spectrometer (LABRAM HR Evolution, Horiba -Jobin-Yvon, France) with the 532 nm excitation of frequency-doubled Nd:YAG laser in the backscattering mode with an 1800 grooves/mm grating. The temperature-dependent Raman studies are carried out using a temperature stage (LINKAM THMS 600, Linkam Scientific Instruments, UK) with the help of a long working distance 50X objective. The power-dependent spectra in ambient temperature conditions are collected with a 100X objective (NA = 0.9) by engaging discrete neutral density (ND) filters. For each ND filter, the final power was measured at the microscope objective end with a power meter (Thorlabs, USA).

Dense silicon NWs growth with a diameter varying between 100 – 500 nm with a large length (of several microns) can be easily noticed from figure 1 (a). Raman spectra are recorded on a bunch of NWs and a strong first-order Raman peak of silicon is observed as shown in figure 1 (b). It may be noted that the width of the Raman spectra is wider than bulk silicon. This NWs from the substrate are then transferred to a carbon-coated copper TEM grid. Once transferred, the grid is inspected for isolated single freestanding NW. This isolated individual NW can sometimes found to be suspended on ruptured carbon film where one end rests on the carbon film and another end lies on the copper grid. To minimize the role of contact on thermal conductivity we consider a suspended single NW of length much larger than the diameter of the laser spot. Figure 2 (a) shows an SEM image of such a freestanding silicon NW of diameter ~112 nm with a suspended length of ~7 μm. A careful analysis of the freestanding isolated NW indicates that the diameter of the NW is not uniform throughout its length i.e. from one side (right side) its diameter gradually decreases. Such tapering of diameter is expected for CVD grown large length NWs. However, for the present study for thermal conductivity measurement, the diameter can be approximated to be uniform. The suspended NW, laser heating are schematically shown on the left side of figure-2(b).

We also schematically present the effective cross-sectional area of NW exposed to laser on the right side of figure-2(b). It can be clearly seen that the diameter of the laser spot is much larger as compared to the diameter of the suspended NW (details of which will be explained in a later section). The Raman spectra are collected from 77 K to 298 K. The laser power is kept low in order to avoid local heating. The acquired spectra are fitted with Lorentzian lineshape to obtain the Raman peak positions and peak width (FWHM).

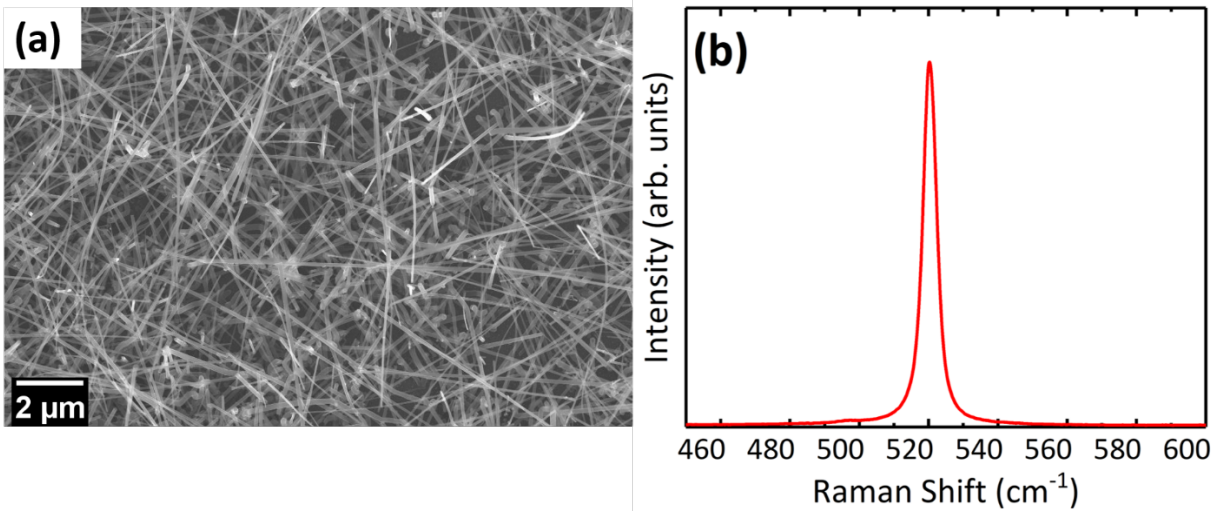

**Figure-1: (a) SEM image of silicon NWs (b) Corresponding room temperature Raman spectra.**

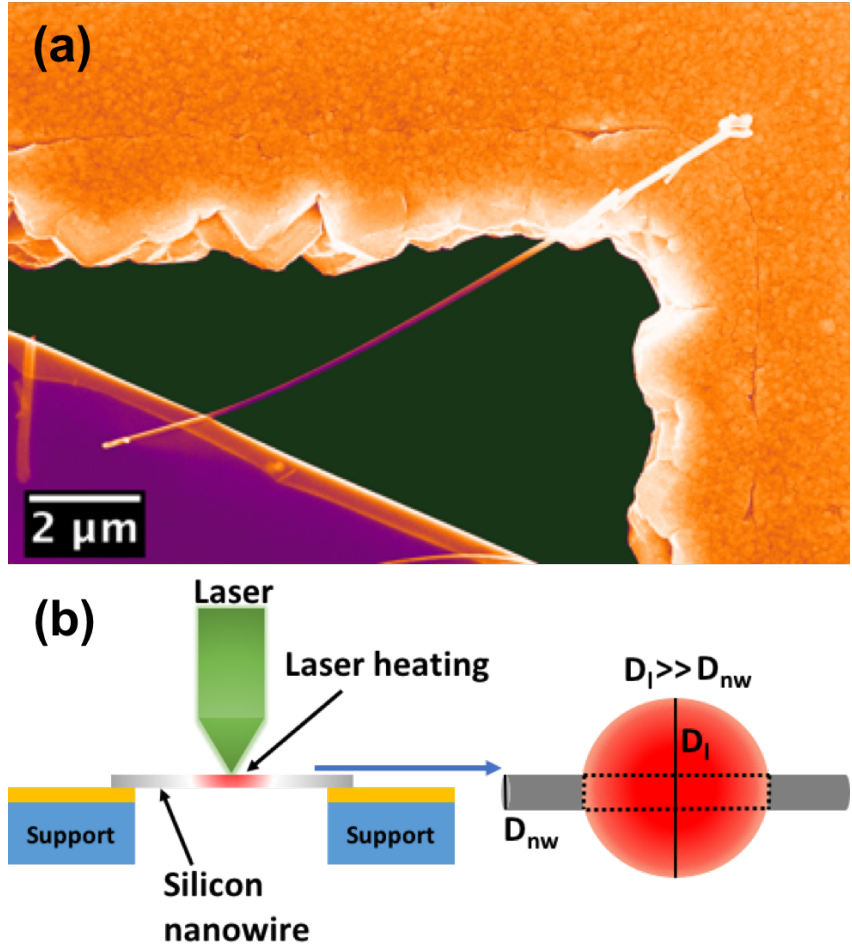

**Figure-2:(a)** False-colored SEM image of a freestanding single silicon NW on TEM grid **(b)** Schematic diagram of free-standing single silicon NW (left) and comparison of the dimension of silicon NW, laser spot size (right). The dotted rectangular section shows an effective cross-sectional area of silicon NW exposed to the laser.

**RESULTS AND DISCUSSION:**

Thermal conductivity measurement using Raman spectroscopy would result in an inaccurate estimation for a sample directly placed on a substrate. This is due to the fact that as a part of the heat will dissipate to the substrate. In view of this most of the measurements have been carried out on a free-standing sample.[18,19,22] As reported thermal conductivity estimation by Raman scattering requires a temperature and laser power-dependent studies. First, we carried out a temperature-dependent Raman study. The temperature of the sample is varied from 77 to 298 K. It is worth arguing that high-temperature studies are deliberately eliminated in the present case because at high-temperature silicon samples could undergo oxidation resulting in a core-shell structure of Si/SiO$_2$. This can affect the temperature-dependent studies. Figure 3(a) shows the temperature-dependent Raman spectra. A systematic redshift of the first-order Raman peak position is noticed for the freestanding silicon NW with increasing temperature. This redshift is due to both the volume and temperature contribution as a consequence of anharmonicity of lattice vibration. The Raman peaks are well fitted with the Lorentzian lineshape and the extracted Raman peak position with the temperature are plotted in figure 3(b). The peak position as a function of temperature shows a linear relationship as fitted using $\omega = \omega_0 + \alpha.T$, where $\omega_0$ is the frequency at T= 300 K, α is the first-order temperature coefficient.[18,22] The slope of the curve gives the value of the first-order temperature coefficient (α) and is found to be -0.01776 cm$^{-1}$ K$^{-1}$. Doerk *et al.* [30] have studied the variation of first-order temperature coefficient as a function of the diameter of silicon NW ranging from 20 to 200 nm. The diameter dependent first-order temperature coefficient does not change significantly and our observed first-order temperature coefficient is also consistent with their reported result .[30]

The broadening of the linewidth which is the measurer of phonon lifetime can be explained using the decay of phonon into two phonons or three phonons processes satisfying the conservation law.[31,32] Considering a three phonons process, the line width can be expressed as,

$$\Gamma(T) = \Gamma_0 + E\left[1 + \frac{2}{e^x - 1}\right] + F\left[1 + \frac{3}{e^y - 1} + \frac{3}{(e^y - 1)^2}\right] \quad (1)$$

where $x = \frac{\hbar\omega}{2.k_B.T}$ , $y = \frac{\hbar\omega}{3.k_B.T}$ , $\Gamma_0$ is FWHM of the silicon NW in low temperature and *E, F* is cubic and quartic anharmonic constants respectively.[33] The variation of FWHM as a function of temperature is well fitted by the above equation for three phonon decay (Fig. 3 (b)) with cubic and quartic anharmonic constants of -0.3097 and 0.4375 respectively.

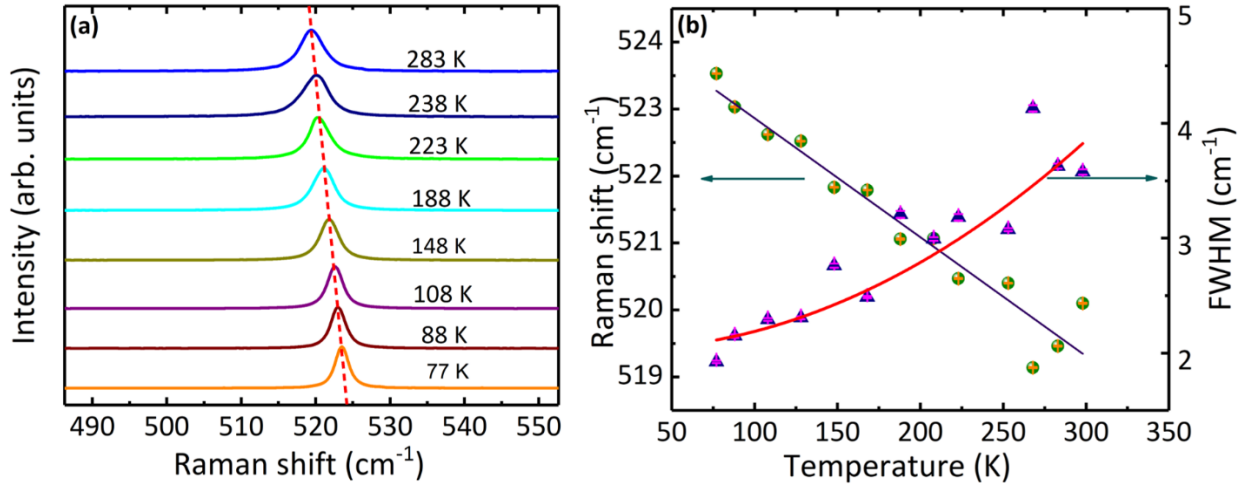

**Figure-3:** (a) First-order Raman spectra of freestanding silicon NW as a function of temperature. The dotted line is the guide to the eye. (b) Variation of Raman peak position and FWHM with temperature. Filled dots and filled triangles represent the peak position and FWHM, respectively. Solid lines are the fits to the experimental data.

Laser power dependent Raman studies with the variation of power is analogous to the temperature-dependent Raman spectra, it signifies that a strong laser power on the sample results in the localized heating.[19] The thermal measurement requires a careful determination of the diameter of the laser beam spot ($D_l$) and actual power absorbed by the NW. A theoretical estimation of the diffraction-limited spot size ($D_l = 2\lambda/\pi NA$) puts the lower limit on the spot size to be 0.38μm where λ is the wavelength of the incident light (532 nm), NA is the numerical aperture of the objective lens (0.9), respectively. In our case, we assume the spot size to be ~ 0.5 μm which is much larger than the NW diameter (112 nm). As the laser beam spot is much larger than the silicon NW, the power absorbed by the silicon NW is not the same as the incident laser power. During the experiment, the NW was placed carefully under the laser spot such that the diameter of the laser ($D_l$) spot coincides with the NW, as schematically shown in figure 2 (b). Under such geometry, the cross-sectional area of the silicon NW illuminated by the laser is $D_l \cdot D_{nw}$, where $D_{nw}$ is the diameter of the silicon NW. As mentioned our laser spot size is much larger than the diameter of silicon NW, we followed the effective power calculation of Doerk *et al.*[23] The power absorbed ($P_a$) by the silicon NW is $\frac{4.D_{nw}.D_l}{\pi.D_l^2}$ times the laser power ($P_l$) and the absoptivity ($A$) and can be expressed as follows.

$$P_a = \frac{4.D_{nw}.D_l}{\pi.D_l^2} A.P_l \qquad (2)$$

It may be noted here that the actual laser power as measured under the objective lens is varied from 43 µW to 1.2 mW during the experiment. The absorptivity ($A$) can be deduced using $A = 1 - \frac{(n-1)^2+k^2}{(n+1)^2+k^2}$, where $n$ is the refractive index and $k$ is the extinction coefficient.[34] For the laser wavelength of 532 nm, the values of $n$ and $k$ are about 4.15 and 0.052 for bulk silicon[35] giving the absorptivity ($A$) of the silicon NW to ~ 62 %. Using the above formula, the effective power $P_a$ is calculated to be in the range of 7 µW to 201 µW. Figure 4 (a) shows the Raman spectra of the freestanding NW as a function of effective laser power.

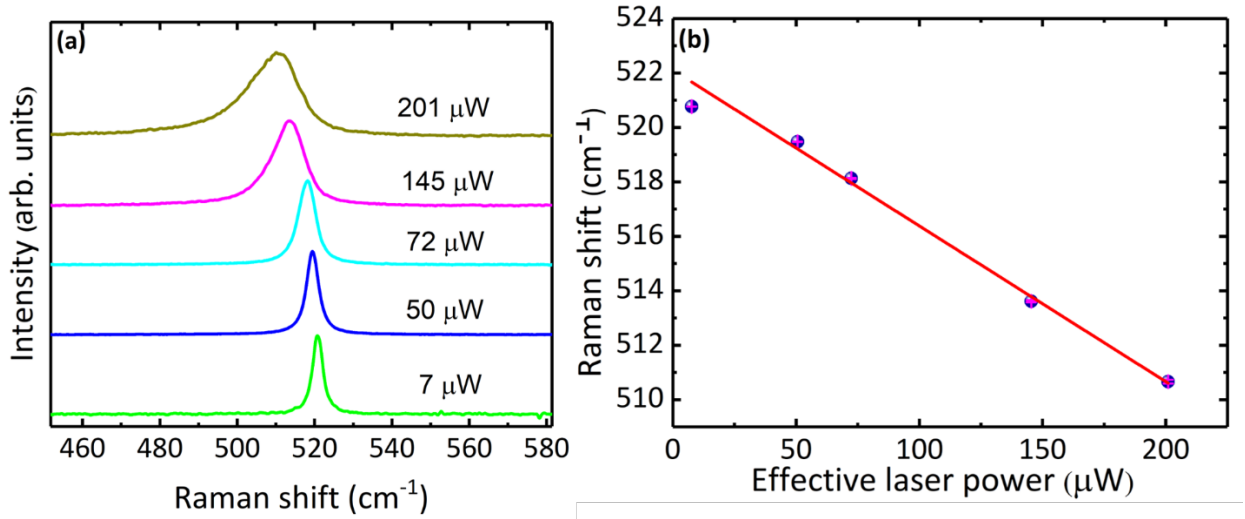

**Figure-4: (a) Raman Spectra of the free-standing NW for different absorbed laser powers (b) Variation of Raman peak position with effective laser power. The solid line is a linear fit**.

The symmetrical broadening of the Raman line and shifting of the first-order Raman peak position towards the lower wave-number side is also observed. The variation of Raman peak position with different power is shown in the Figure-4(b) and is fitted by a straight line and the slope ($\frac{\delta\omega}{\delta P_a}$) of $\omega \sim P_a$ is found to be -0.05721 cm$^{-1}$ (μW)$^{-1}$ from the linear fit.

Here, we would like to emphasize that silicon is always associated with a native oxide layer irrespective of its synthesis methods. The thickness of the oxide layer can vary depending on process conditions. The role of the oxide layer is very important in determining various properties of silicon NW. For an instant, during the electrical transport measurement of isolated silicon NW, the native oxide layer is removed carefully before making the electrical contacts. Being an insulator and highly dielectric material, the presence of silicon dioxide (SiO$_2$) layer on silicon NW can significantly affect its thermal properties. It has been experimentally observed that one requires higher laser power to get a significant Raman shift in silicon NW having a thicker oxide layer analogous to the thin oxide layer covered silicon NW. The work by Piscanec et al.[36], Zixue Su et al.[33] shows shifts in the Raman spectra for both thin (~2 nm) and thick (~10 nm) oxide layers covering silicon NW respectively. In our case, significant changes in power-dependent Raman spectra are indicative of the existence of a very thin (~2-3 nm) oxide layer covering the silicon NW.

Thermal conductivity of a sample can be determined by considering the heat distribution i.e. (i) radially heat distribution from the middle position of the sample towards the heatsink if the diameter of the sample is larger than the diameter of the laser spot (ii) longitudinal heat distribution from the middle position of the sample towards the heatsink if the diameter of the sample is comparable to the diameter of the laser spot.[18] Transport of heat through the surface of a sample with a cross-sectional area $S$ is formulated as,

$$\frac{\partial Q}{\partial t} = -\kappa . \oint \nabla T . dS \qquad (3)$$

where κ is the thermal conductivity of the sample, $\frac{\partial Q}{\partial t}$ is the rate of heat transferred through the sample, $\nabla T$ is the temperature gradient. When the diameter of the sample is comparable to the diameter of the laser spot, the thermal conductivity of the sample can be formulated from the equation (3) as $\kappa = \frac{L}{2.S} . \frac{\Delta P_l}{\Delta T}$, where L is the distance from the laser-heated position of the sample to the heat sink and $\Delta T$ is the change in temperature due to the variation of the laser power. [18] The value of L is about ~3.4 µm as measured from the SEM image [figure 2(a)]. We note that in our case the diameter of the laser spot is quite larger than the diameter of the silicon NW. However, if we consider the portion of the laser power ($P_a$) absorbed by the silicon NW as shown in figure 2 (b), then the heat distribution will be longitudinally allowing us to use the formula: $\kappa = \frac{L}{2.S} \cdot \frac{\Delta P_a}{\Delta T}$. We assumed that the increase in the laser power i.e. increasing the local temperature of the silicon NW does not affect the temperature of the heat sink and the local heat generated is transferred uniformly to the heat sink. With the help of linear fit of the ω ~ T graph i.e. $\omega = \omega_0 + \alpha.T$, the above formula can be modified to $\kappa = \frac{4.L}{2.\pi d^2} \cdot \alpha \cdot (\frac{\delta \omega}{\delta P_a})^{-1}$.

With laser spot diameter 0.5 µm, the diameter of silicon NW 112 nm, length of silicon NW ~3.4 µm, the value of first-order temperature coefficient ($\alpha$) as - 0.01776 cm$^{-1}$K$^{-1}$ and value of $\frac{\delta \omega}{\delta P_a}$ as - 0.05721 cm$^{-1}$(µW)$^{-1}$, the thermal conductivity of the single silicon NW sample is estimated to be ~ 53 W/m.K.

Doerk *et al.* reported the variation of thermal conductivity with NW diameter and for the silicon NW with diameter ~ 115 nm, room temperature thermal conductivity is about 40 W/m.K.[23] Similar thermal conductivity values were also reported by Li *et al.* for individual NWs with a nearly similar diameter.[14] Although our thermal conductivity estimation is in broad agreement with that of Doerk *et al.*[23] and Li *et al.*[14], our estimated value is slightly higher. Thermal conductivity of a material is highly dependent on the external strain which affects the phonon group velocity as well as specific heat. Equilibrium Molecular Dynamics calculations by Li *et al.* on thermal conductivity of silicon NWs as a function of strain reveal higher thermal conductivity under compressive strain and lower thermal conductivity under tensile strain as compared to the unstrained one.[37–39] In addition, surface roughness causes more boundary scattering which greatly reduces the thermal conductivity of the silicon NW.[40] Therefore, the small discrepancy in our thermal conductivity could be due to the existence of compressive strain on the silicon NW. The suppressed thermal conductivity of silicon NW as compared to the bulk silicon ($\kappa$ = 148 W/m.K)[41] is due to increased phonon boundary scattering.

**CONCLUSIONS:** In summary, Raman spectroscopy is employed to determine the thermal conductivity of the CVD grown free-standing silicon NW of diameter 112nm. Free-standing silicon NWs are obtained by transferring the NWs from the substrate to a carbon-coated copper TEM grid. Isolated individual NW can sometimes found to be suspended on ruptured carbon film where one end rests on the carbon film and another end lies on the copper grid. To minimize the role of contact on thermal conductivity we consider a suspended single NW of length much larger than the diameter of the laser spot. The temperature dependence of Raman spectra provides the first-order temperature coefficient($\alpha$) which is -0.01776 cm$^{-1}$ K$^{-1}$ and laser-induced Raman spectra undergo a redshift which provides a slope of -0.05721 cm$^{-1}$ ($\mu$W)$^{-1}$. By means of the first-order temperature coefficient ($\alpha$) value and slope of the Raman peak position~ power graph, the thermal conductivity of the free-standing single silicon NW is found to be ~ 53 W/m.K.

**ACKNOWLEDGEMENTS**:

We acknowledge P. V. Satyam for extending the SEM facility and Puspendu Guha for assistance during the experiment.